\documentstyle[psfig,mncite,subfigure]{mn}

\newcommand{\rmsub}[2]{#1_{\rm #2}}

\begin{document}

\title{A new upper limit on the reflected starlight from $\tau$ Bootis $b$}
\author
[C.Leigh, A.Collier~Cameron, K.Horne, A.Penny and D.James]
{\Large Christopher Leigh$^{1}$, Andrew Collier Cameron$^{1}$, Keith Horne$^{1}$,
Alan Penny$^{2}$ and David James$^{3}$\\
\small $^{1}$ University of St Andrews, St Andrews, Fife, KY16 9SS, U.K \\
$^{2}$ Rutherford Appleton Laboratory, Chilton, Didcot, Oxon, OX11 0QX, U.K \\
$^{3}$ Observatoire de Grenoble, F-38041 Grenoble, Cedex 9, France}
\date{14 June 2003}

\maketitle

\begin{abstract}
Using improved doppler tomographic signal-analysis techniques we have
carried out a deep search for starlight reflected from the giant
planet orbiting the star $\tau$ Bootis. We combined echelle spectra
secured at the 4.2 m William Herschel telescope in 1998 and 1999
(which yielded a tentative detection of a reflected starlight
component from the orbiting companion) with new data obtained in 2000
(which failed to confirm the detection). The combined dataset
comprises 893 high resolution spectra with a total integration time of
$75^{hr} 32^{min}$ spanning 17 nights. We establish an upper limit 
on the planet's geometric albedo $p<0.39$ (at the 99.9\% significance 
level) at the most probable orbital inclination $i\simeq 36^\circ$,
assuming a grey albedo, a Venus-like phase function and a
planetary radius $R_{p}=1.2 R_{Jup}$. We are able to rule out some
combinations of the predicted planetary radius and atmospheric albedo
models with high, reflective cloud decks. Although a weak candidate
signal appears near to the most probable radial velocity amplitude, its
statistical significance is insufficient for us to claim a detection
with any confidence.
\\
\\
\\
\end{abstract}
\section{Introduction}
\label{sec:intro}
Two years after the discovery of a planetary companion to 51 Pegasi by
\scite{mayor95} came the identification of a similar object
orbiting the F7V star $\tau$ Bootis \cite{butler97}. Among the more
surprising features of these discoveries were their short orbital
periods, putting both objects very close to their parent stars. 
Indeed, of the 102 extra-solar planets currently known, 19 reside
within 0.1 AU of the parent star. In addition to posing many
theoretical questions as to how they came to be there, the existence
of these objects in such close orbits does open the possibility of
their detection through the reflection of the host stellar spectrum.

Shortly after the $\tau$ Bootis b detection, two teams
\cite{charb99,cameron99} initiated a spectroscopic search for the
reflected light component of the orbiting planet. Charbonneau et al. 
conducted 3 nights observation in March 1997 using the HIRES echelle
spectrograph mounted on the Keck I 10m telescope on Mauna Kea, Hawaii. 
Although the restricted spectral range 465.8 - 498.7 nm provided a non
detection, the signal-to-noise ratio ($SNR \sim 1500$) was sufficient
to impose a relative reflected flux limit $f_p/f_* < 5 \times
10^{-5}$, assuming a grey albedo reflection of the stellar spectrum. 
This implies a geometric albedo limit $p < 0.3$ over the spectral
range investigated, assuming a planetary radius of $1.2~R_{Jupiter}$. 
Cameron et al.  obtained 10 nights of data during 1998 and 1999, with
the Utrecht echelle spectrograph (UES) on the 4.2 m William Herschel
telescope. By using a least-squares deconvolution (LSD) technique
\cite{donati97} on $\sim 2300$ spectral lines in the range 385 - 611
nm, they identified a {\it probable} reflected-light feature with
$f_p/f_* \sim 7.8 \times 10^{-5}$ \cite{cameron99}. This detection
indicated an orbital velocity amplitude $K_p = 74 \pm 3$~km~s$^{-1}$
which, when combined with the planet's orbital velocity $V_p =
152$~km~s$^{-1}$, suggested an inclination for the system $i = 29^o$ and
a radius $R_{p} = 1.8 R_{J}$, assuming a grey geometric albedo $p =
0.55$. A bootstrap Monte Carlo analysis gave a 5\% probability that
the feature was an artefact of noise. Subsequent observations over 7
nights in March-May 2000, however, failed to confirm the detection
\cite{cameron2001}.

\begin{table*}
\caption{System parameters for $\tau$ Bootis and its planetary companion}
\begin{tabular}{ccc}
\hline \hline
Parameter & Value (Uncertainty) & References \\
\hline
$\bf{Star:}$ \\
Spectral Type		& F7V 		& \scite{fuhrmann98,gonzalez98} \\
$M_{V}$			& 4.496 (0.008)	& \scite{fuhrmann98,gonzalez98} \\
Distance (pc)	& 15.6 (0.17)	& \scite{perryman97} \\
$T_{Eff}$ (K)		& 6360 (80)	& \scite{fuhrmann98,gonzalez98} \\
$M_{*} (M_\odot)$	& 1.42 (0.05)	& \scite{fuhrmann98,gonzalez98} \\
$R_{*} (R_\odot)$	& 1.48 (0.05)	& \scite{fuhrmann98,gonzalez98} \\
$\left[Fe/H\right]$	& 0.27 (0.08)	& \scite{fuhrmann98,gonzalez98} \\
v$\sin i$ (km~s$^{-1}$)	& 14.9 (0.5)	& \scite{henry2000} \\
Age (Gyr)		& 1.0 (0.6)	& \scite{fuhrmann98} \\
\hline
$\bf{Planet:}$ \\
Orbital Period $P_{orb}$ (days)	& 3.31245 (0.00003)	& Marcy, private communication \\
Transit Epoch $T_{0}$ (JD)	& 2451653.968 (0.015)	& Marcy, private communication \\
$K_{*} (ms^{-1})$		& 469 (5)		& \scite{butler97} \\
a (AU)				& 0.0489		& \scite{butler97}(revised for this paper) \\
$M_{P}\sin i (M_{Jup})$		& 4.38			& \scite{butler97}(revised for this paper) \\
\hline \hline \\
\end{tabular}
\label{parameters}
\end{table*}

Here we report the results of a new, deep search for the reflected
light signal via a full re-analysis of the WHT data, combining all 17
nights of echelle spectra obtained during the 1998, 1999 and 2000
observing seasons. In Section~\ref{sec:parameters} we use the
measured system parameters to determine prior probabilities for the
planet's orbital velocity amplitude and the fraction of the star's
light that it intercepts. In Section~\ref{sec:obser} and
\ref{sec:extraction} we describe the acquisition and extraction of the
echelle spectra. Section~\ref{sec:aldecon} details the methods used
to extract the planetary signal from the data, by combining the
profiles of thousands of stellar absorption lines recorded in each
echellogram into a time series on which we use a matched-filter method
to measure the strength of the reflected-light signal. 
Section~\ref{sec:simulated} describes how we test and calibrate the
analysis through the addition of a simulated planetary signal. 
Finally Section~\ref{sec:results} makes theoretical assumptions about
the size and atmospheric composition of $\tau$ Bootis b in order to
place upper limits on its respective geometric albedo and radius. We
also discuss the plausibility of a candidate reflected-light signature
that appears in the data close to the most probable velocity amplitude
and signal strength.

\section{System Parameters}
\label{sec:parameters}
$\tau$ Bootis (HD 120136, HR 5185) is a late-F main sequence star
with parameters as listed in Table~\ref{parameters}. High precision
radial velocity measurements over a period of 9 years were used to
identify a planetary companion \cite{butler97} whose properties
(as determined directly from the radial velocity studies or
inferred using the estimated stellar parameters) are also
summarised in Table~\ref{parameters}.

The equations detailed below represent a summary of the more comprehensive
derivations described in previous work \cite{cameron99,charb99,cameron02}.

As the planet orbits its host star, some of the starlight incident upon its
surface is reflected towards us, producing a detectable signature within the
observed spectra of the star. This signature takes the form of faint copies
of the stellar absorption lines, Doppler shifted due to the planet's orbital
motion and greatly reduced in intensity ($\sim 10^{-4}$) due to the small
fraction of starlight the planet intercepts and reflects back into space.

With our knowledge of stellar mass and the planet's orbital period we
can estimate the orbital velocity of the planet $V_{p}$. The apparent
radial velocity amplitude $K_{p}$ of the reflected light is given by:
\begin{equation}
K_{p} = V_{p} \sin i = 161 \sin i~\mbox{km~s}^{-1}
\label{eq:kp}
\end{equation}
where the orbital inclination $i$ is, according to the usual convention, the
angle between the orbital angular momentum vector and the line of sight.

For all but the lowest inclinations, the orbital velocity amplitude of the
planet is substantially greater than the broadened widths of the photospheric
absorption lines of the star. Hence, lines in the reflected-light spectrum of
the planet should be Doppler shifted well clear of their stellar counterparts,
allowing a clean spectral separation for most of the orbit.

By isolating the reflected planetary signature, we are in effect observing the 
planet/star flux ratio ($\epsilon$) as a function of orbital phase ($\phi$)
and wavelength ($\lambda$).
\begin{equation}
\epsilon(\alpha,\lambda)\equiv\frac{f_{p}(\alpha,\lambda)}{f_{\star}(\lambda)}
        =p(\lambda)g(\alpha,\lambda)\frac{R_{p}^{2}}{a^{2}}
    = \epsilon_{0}(\lambda)g(\alpha,\lambda)
\label{eq:fluxratio}
\end{equation}
The phase function $g(\alpha,\lambda)$ describes the variation in the
star-planet flux ratio with illumination phase angle $\alpha$. This
is the angle subtended at the planet by the star and the observer, and
varies according to $\cos\alpha = - \sin i .\cos\phi$.

The observations measure $\epsilon(\alpha,\lambda)$ over some range of
orbital phases $\phi$ and hence phase angles $\alpha$. However, the
signal-to-noise ratio and orbital phase coverage of the observations
is not yet adequate to define the shape of the phase function. 
Accordingly, current practice is to adopt a specific phase function
in order to express the results in terms of the
planet/star flux ratio that would be seen at phase angle zero:
\begin{equation}
\epsilon_{0}(\lambda)=p(\lambda)\frac{R_{p}^{2}}{a^{2}}
\label{eq:eps0}
\end{equation}
where $p(\lambda)$ is the wavelength dependent geometric albedo.
Since $a$ is tightly constrained by Kepler's third law to
$a = 0.0489 \left(M_{*}/1.42 M_{\odot}\right)~AU$, the measurements of
$\epsilon_{0}(\lambda)$ measure the product $R_{p}\sqrt{p(\lambda)}$.

While the phase function of a Lambert sphere might be the simplest
form to adopt for $g(\alpha,\lambda)$, we prefer to adopt a phase
function that resembles those for the cloud-covered surfaces of
planets in our own solar system. Jupiter and Venus appear to have
phase functions that are more strongly back-scattering than a Lambert
sphere. Photometric studies of Jupiter at large phase angles from the
{\em Pioneer} and subsequent missions have shown \cite{hovenier89}
that the phase function for Jupiter is very similar to that of Venus,
despite their very different cloud compositions. As a plausible
alternative to the Lambert-sphere formulation, we use a polynomial
approximation to the empirically determined phase function for Venus
\cite{hilton92}. The phase-dependent correction to the planet's
visual magnitude is approximated by:
\begin{equation}
\Delta m(\alpha)=0.09(\alpha/100^\circ) + 2.39(\alpha/100^\circ)^2
- 0.65(\alpha/100^\circ)^3
\label{eq:dm_venus}
\end{equation}
so that
\begin{equation}
g(\alpha)=10^{-0.4 \Delta m(\alpha)}
\label{eq:g_venus}
\end{equation}

In the event of any planetary detection, a careful analysis
of the data should allow us to determine the following information:

(i) $K_{P}$, the planet's projected orbital velocity, from which we
obtain the orbital inclination of the system and hence the mass of
the planet, since $M_{P}\sin i$ is known from the star's Doppler wobble.

(ii) $\epsilon_{0}$, the maximum flux ratio observed, with which we
can constrain the planet's radius since $\epsilon_{0} = p(R_{P}/a)^{2}$,
where $p$ is the geometric albedo of the planetary atmosphere.
Alternatively we can adopt a theoretical radius to constrain the albedo
of a given atmospheric model.

\subsection{Rotational broadening}
\label{sec:rotational}

The rotational broadening of the direct starlight and chromospheric Ca
II H \& K emission flux suggest that the star's rotation is 
synchronised with the orbit of the planet \cite{baliunas97,henry2000}. 
In a tidally locked system there is no relative motion between the
surface of the planet and the surface of the star, so the planet will
reflect a non-rotationally broadened stellar spectrum, with typical
line widths dominated by turbulent velocity fields in the stellar
photosphere. These motions were estimated by \scite{baliunas97} to be
of the order $\sim 4$~km~s$^{-1}$. Any absorption lines attributed to
the planet's atmosphere are thus likely to be much narrower than the
stellar lines.

\subsection{Orbital Inclination}
\label{sec:inclination}

In the first instance we can rule out inclinations $i>80^{\circ}$ due
to the absence of transits in high-precision photometry
\cite{henry2000}. Furthermore, if we assume the star's rotation to be
tidally locked to the planet's orbit, we can use the projected
equatorial rotation speed of the host star $v\sin i = 14.9 \pm 0.5$ km
s$^{-1}$ \cite{henry2000} to loosely constrain the orbital inclination
to $i \sim 40^\circ$. We would thus expect a projected orbital velocity
amplitude close to $K_{p} \sim 100$~km~s$^{-1}$.

\subsection{Planet Radius}
\label{sec:radius}

The HD 209458b transit detection of \scite{charb2000} provided the
first confirmation of the gas giant nature of close-in extra solar
planets, and yielded a radius in good agreement with the predictions
of past and current interior structure models
\cite{guillot96,seager98,marley99,burrows2000,seager2000}. In short,
the planet radius evolves with time and depends on the planet mass. 
For our purposes, this defines a range of theoretically plausible
radii at each possible value of the planet mass. The range of
possible planet radii was computed specifically for $\tau$ Bootis b by
\scite{burrows2000}, allowing for uncertainties in the orbital
inclination and hence the planet's mass. Their radiative-convective
gas giant models predict upper limits on the planet's radius of 1.58
$R_{Jup}$ for $M_{p}$ = 7 $M_{Jup}$, and 1.48 $R_{Jup}$ for $M_{p}$ =
10 $M_{Jup}$.

\subsection{Prior Estimates of System Parameters}
\label{sec:prior}

\begin{figure}
\psfig{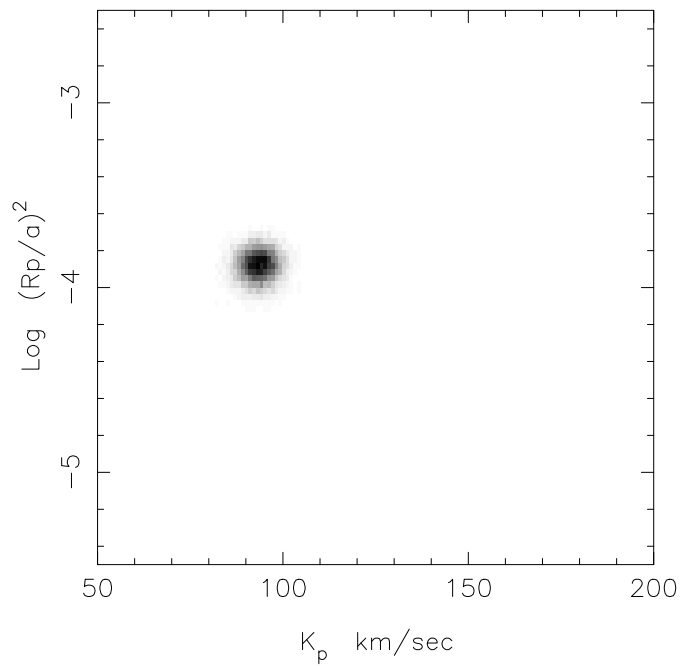}
\caption[]{The greyscale shows the prior joint probability density
function (PDF) for projected orbital velocity $K_{p}$ and the squared
ratio $(R_{p}/a)^{2}$ of the planet radius to the orbit radius, based
on the measured system parameters. Darker shades in the greyscale
denote greater probabilities of the planet having the corresponding
combination of $(R_{p}/a)^{2}$ and $K_{p}$. The PDF shows that the
planet is most likely to have $K_{p} \simeq 94$~km~s$^{-1}$ and
$(R_{p}/a)^{2}\simeq 1.44 \times 10^{-4}$.}
\label{fig:prior}
\end{figure}

\begin{table*}
\caption{Journal of observations. The UTC mid-times and orbital
phases are shown for the first and last spectral exposures
secured on each night of observation. The number of exposures is
given in the final column.}
\label{tab:journal}
\begin{tabular}{cccccc}
\\
UTC start             & Phase   &  UTC End             & Phase  & Exposures\\
                      &         &                      &        &          \\
1998 Apr 09 22:09:43 & 0.425 & 1998 Apr 10 05:37:05 & 0.519 & 107 \\
1998 Apr 10 22:04:40 & 0.726 & 1998 Apr 11 06:20:54 & 0.830 & 113 \\
1998 Apr 11 22:09:28 & 0.029 & 1998 Apr 12 05:58:44 & 0.127 &  81 \\
1998 Apr 13 23:02:54 & 0.644 & 1998 Apr 14 05:37:05 & 0.733 &  33 \\
                     &       &                      &       &     \\
1999 Apr 02 22:06:45 & 0.493 & 1999 Apr 03 06:08:47 & 0.594 &  25 \\
1999 Apr 25 21:43:17 & 0.441 & 1999 Apr 26 05:31:59 & 0.539 &  40 \\
1999 May 05 21:56:59 & 0.463 & 1999 May 06 04:51:57 & 0.550 &  60 \\
1999 May 25 20:59:45 & 0.488 & 1999 May 26 03:56:18 & 0.576 &  51 \\
1999 May 28 20:52:23 & 0.390 & 1999 May 29 03:04:35 & 0.467 &  47 \\
1999 Jun 04 20:22:34 & 0.497 & 1999 Jun 05 00:07:23 & 0.548 &  23 \\
                     &       &                      &       &     \\
2000 Mar 14 23:14:49 & 0.271 & 2000 Mar 15 06:54:52 & 0.366 &  48 \\
2000 Mar 15 22:51:55 & 0.567 & 2000 Mar 16 06:49:04 & 0.669 &  45 \\
2000 Mar 24 22:21:55 & 0.278 & 2000 Mar 25 06:47:50 & 0.385 &  34 \\
2000 Apr 23 20:38:59 & 0.311 & 2000 Apr 24 05:07:31 & 0.420 &  45 \\
2000 Apr 24 20:58:32 & 0.619 & 2000 Apr 25 05:14:11 & 0.724 &  57 \\
2000 May 13 20:47:01 & 0.349 & 2000 May 14 03:12:08 & 0.434 &  41 \\
2000 May 17 20:24:14 & 0.557 & 2000 May 18 03:52:37 & 0.649 &  43 \\
\\
\end{tabular}
\end{table*}

In searching for a faint reflected-light signature from a planet with
an unknown orbital inclination, it is useful to know in advance how
the planet's observable properties ought to depend on the orbital
inclination. We do this by constructing the {\it a priori}
probability density functions for the various observable properties of
the planet described in Table~\ref{parameters}. This helps us to
determine whether any faint candidate reflection signature is physically
plausible, given our existing knowledge of the system's parameters. 
We do not want to be guided too closely by theory, but values outside
the plausible ranges would pose difficulties for current thinking.

Fig. \ref{fig:prior} shows the probability distributions for the
planet's radial velocity amplitude $K_{p}$ and the 
quantity $(R_{p}/a)^{2}$, based on a Monte Carlo simulation
using the expressions given in Section~\ref{sec:parameters}.  We
assume Gaussian distributions for the measured stellar mass ($1.42 \pm
0.05 M_\odot$), radius ($1.48 \pm 0.05 R_\odot$), stellar reflex
velocity ($469 \pm 5$ m~s$^{-1}$) and the radius of the planet. The
planet radius and uncertainty range were obtained from the theoretical
mass-radius relations \cite{guillot96,burrows2000} described above. 
Both works find the most probable radius to be $1.2 \pm 0.1 R_{Jup}$,
assuming $\tau$ Bootis b to have an age of 1 Gyr.

Furthermore, we assume the star's rotation is tidally locked to the
planet's orbit. We thus generate a distribution of $\sin i$ values
based on Gaussian distributions for the projected stellar rotation
speed ($v\sin i = 14.9\pm 0.5$~km~s$^{-1}$) and a stellar rotation
period (3.3 $\pm$ 0.1 days) closely bound to the orbital period of the
planet. A further restriction applies where the tidal synchronisation
timescale for the primary's rotation,
\begin{equation}
\tau_{sync}\simeq1.2\left(\frac{M_{p}}{M_{*}}\right)^{-2}\left(\frac{a}{R_{*}}\right)^{6}~years
\label{eq:timesynch}
\end{equation}
is longer than the main-sequence liftime of the host star
$\tau_{ms}\simeq10^{10}(M_{*}/M_{\odot})^{-3}$ years, in which case we
reject that model from the Monte Carlo analysis.

The resulting probability map (Fig.~\ref{fig:prior}) shows a $K_{p}$
distribution centred on $\simeq 94$~km~s$^{-1}$, with the most likely
value of $(R_{p}/a)^{2}~\simeq 1.44\times 10^{-4}$. The projection of
this PDF on to the orbital velocity axis defines the region of
parameter space in which we can be confident that a detection would
occur if it were present in the data, given our prior knowledge of the
system parameters. We use the $K_{p}$ projection of the PDF in the
subsequent analysis to test the plausibility of any candidate features
which appear in the data, by modifying the posterior probability
distribution to assess the false alarm probability (see
Section~\ref{sec:grey}). Secondly, our upper limits on $\epsilon_{0}$
are sensitive to the orbital inclination, so we adopt the most
probable $K_{p}$ in order to determine the most plausible upper limits
on the planet's radius and albedo.

For any given albedo model we can also use the data to determine upper
limits on $(R_{p}/a)^{2}$ instead of the opposition flux ratio
$\epsilon_{0}$. The projection of the PDF onto $(R_{p}/a)^{2}$ thus
allows us to compare the effective reflection area of the planet
directly with model predictions.  Unlike the projection on to $K_{p}$,
however, the prior probability distribution for $(R_{p}/a)^{2}$ plays
no role in assessing the plausibility or otherwise of a candidate
detection.

\section{Observations}
\label{sec:obser}

We observed $\tau$ Bootis during 1998, 1999 and 2000 using the Utrecht
Echelle Spectrograph on the 4.2 m William Herschel Telescope at the
Roque de los Muchachos Observatory on La Palma. The detector was a
single SITe 1 CCD array containing some $2048\times 2048$ 13.5-$\mu$m
pixels. The CCD was centred at 459.6 nm in order 124 of the 31 g
mm$^{-1}$ echelle grating, giving complete wavelength coverage from
407.4 nm to 649.0 nm with minimal vignetting. The average pixel
spacing was close to 3.0~km~s$^{-1}$, and the full width at half
maximum intensity of the thorium-argon arc calibration spectra was 3.5
pixels, giving an effective resolving power $R=53000$.

Table \ref{tab:journal} lists the journal of observations for the 17
nights of data which contribute to the analysis presented in this
paper. In the first year (1998) the stellar spectra were exposed
between 100 and 200 seconds.  For 1999 and 2000, the stellar spectra
were exposed for between 300 and 500 seconds, depending upon seeing,
in order to expose the CCD to a peak count of 40000 ADU per pixel in
the brightest parts of the image.  A 450-s exposure yielded about
$1.2\times 10^{6}$ electrons per pixel step in wavelength in the
brightest orders in typical (1 arcsec) seeing after extraction. We
achieved this with the help of an autoguider procedure, which improves
efficiency in good seeing by trailing the stellar image up and down
the slit by $\pm 2$ arcsec during the exposure to accumulate the
maximum S:N per frame attainable without risk of saturation. Note
that the 450-s exposure time compares favourably with the 53-s readout
time for the SITe 1 CCD in terms of observing efficiency -- the
fraction of the time spent collecting photons is above 90\%. 
Following extraction, the S:N in the continuum of the brightest orders
is typically 1000 per pixel.

\section{Spectrum extraction}
\label{sec:extraction}

One-dimensional spectra were extracted from the CCD frames using an   
automated pipeline reduction system built around the Starlink ECHOMOP
and FIGARO packages. Nightly flat-field frames were summed from 50 to
100 frames taken at the start and end of each night, using an
algorithm that identified and rejected cosmic rays and other
non-repeatable defects by comparing successive frames. Due to physical
movement of the chip mounting between and during observation runs, it was
found that the level of noise was reduced by the use of nightly  
flat fields rather than master flat fields for the entire
year's observations.

The initial tracing of the echelle orders on the CCD frames was 
performed manually on the spectrum of $\tau$ Bootis itself, using 
exposures taken for this purpose without dithering the star up and    
down the slit. The automated extraction procedure then subtracted the
bias from each frame, cropped the frame, determined the form and
location of the stellar profile on each image relative to the trace,
subtracted a linear fit to the scattered-light background across the
spatial profile, and performed an optimal (profile and inverse
variance-weighted) extraction of the orders across the full spatial
extent of the object-plus-sky region. Nightly flat-field balance
factors were applied in the process using the 50 to 100 frames
obtained at the start and end of each night of observations. In all,
55 orders ( orders 88 to 142 ) were extracted from each exposure, giving
full spectral coverage from 407.4 to 649.1 nm with good overlap.

\section{Extracting the planet signal}
\label{sec:aldecon}

For a bright, cloudy model planet with $p=0.4$ and $R_{p}=1.2
R_{Jup}$, we expect the flux of starlight scattered from the planet 
to be no more than one part in 18000 of the flux received directly
from $\tau$ Bootis itself, even at opposition ($\alpha = 0$).  In
order to detect the planet signal, we first subtract the direct
stellar component from the observed spectrum, leaving the planet
signal embedded in the residual noise pattern. A detailed description
of this procedure is given in \scite{cameron02} Appendix A. The planet
signal consists of faint Doppler-shifted copies of each of the stellar
absorption lines. After cleaning up any correlated fixed-pattern
noise remaining in the difference spectra (see \pcite{cameron02}
Appendix B), we then create a composite residual line profile, by
fitting to the thousands of lines recorded in each echellogram
(\pcite{cameron02} Appendix C). Finally we use a matched-filter
analysis (\pcite{cameron02} Appendix D) to search for features in the
time-series of composite residual profiles whose temporal variations
in brightness and radial velocity resemble those of the expected
reflected-light signature. For an assumed albedo spectrum
$p(\lambda)$ and orbital velocity amplitude $K_{p}$, the fit of the
matched filter to the data measures $(R_{p}/a)^{2}$.

\section{Analysis Changes}
\label{changes}
 
In this new analysis of the $\tau$ Bootis data, we
have made the following significant changes to the processing undertaken
for the original \scite{cameron99} paper -
 
(i) The inclusion of the year 2000 data, which adds seven nights' data
taken at optimally-illuminated orbital phases to the analysis.

(ii) Full re-extraction of all three years' data, again using optimal
methods, and providing an increase in the spectral range by two
echelle orders or $\sim$ 15 nm.

(iii) The use of nightly flat-field frames in the extraction
routine, rather than the previous whole year flat-fields. Post
extraction analysis showed a $\sim 4 $\%\ reduction in noise.
 
(iv) Increases in computational processing power over the intervening
two years has allowed the analysis to be conducted on individual
echelle frames, rather than having to co-add the spectra into groups
of four prior to the deconvolution and matched-filter analysis.
 
(v) The inclusion of a Principal Component Analysis routine (PCA), as
detailed in \scite{cameron02} Appendix B, to remove correlated
fixed-pattern noise that was appearing in the difference spectra (i.e.
raw spectra - stellar template frames).

(vi) The use of a more stringent calibration technique, described at
Appendix~\ref{sec:calibration}, to quantify and correct for the
fraction of the planetary signal lost during the stellar subtraction,
deconvolution and PCA routines. With this we produce shallower but
more realistic upper limits than were stated by
\scite{cameron2001}, who assumed no loss of signal.

\section{Simulated planet signatures}
\label{sec:simulated}

We verified that a faint planetary signal is preserved through the
above sequence of operations in the presence of realistic noise
levels, by adding a simulated planetary signal to the observed
spectra. We also use the simulated signal to calibrate the strength
of any detected signal (Appendix~\ref{sec:calibration}). The
simulations were based on the assumption that the planet's rotation is
close to being tidally locked, always keeping the same face towards
the star. The resulting broadening of the spectral lines is therefore
dominated by convective motions on the star's surface, estimated at
$\simeq 4$~km~s$^{-1}$ \cite{baliunas97}. For our simulations we
chose to use the slowly rotating giant star HR 5694, observed on
several nights in 1999.  HR 5694 is a F7III spectral type of similar
temperature and elemental abundance to $\tau$ Bootis, but with an
estimated $v\sin i \simeq 6.4 \pm 1$~km~s$^{-1}$, making it well
suited to represent the reflected starlight \cite{baliunas97}.

\begin{figure}
\psfig{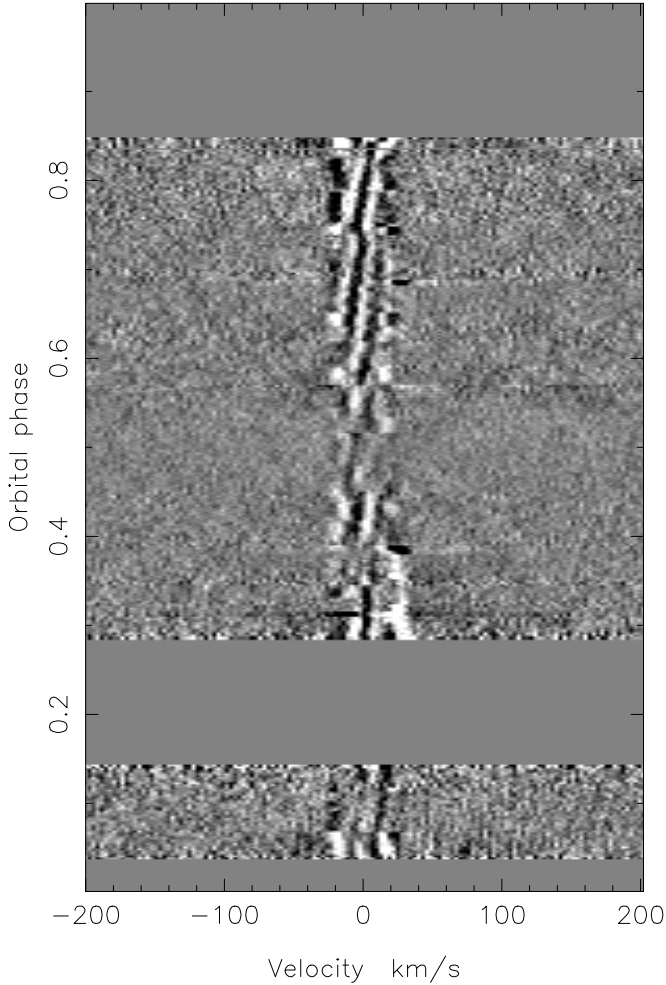}
\caption[]{ Time series of deconvolved profiles derived from the
original WHT spectra, and secured over 17 nights observations in 1998,
1999 and 2000, but with the addition of a simulated planet signal at
an inclination of $60^\circ$. The injected signal is that of a planet
with geometric albedo $p=0.5$ and radius $1.4 R_{Jup}$. The planetary
signature appears as a dark sinusoidal feature crossing from right to
left as phase increases and centred on the superior conjunction at
phase 0.5.}
\label{fig:ph_a0.5r1.4ax60}
\end{figure}
 
\begin{figure}
\psfig{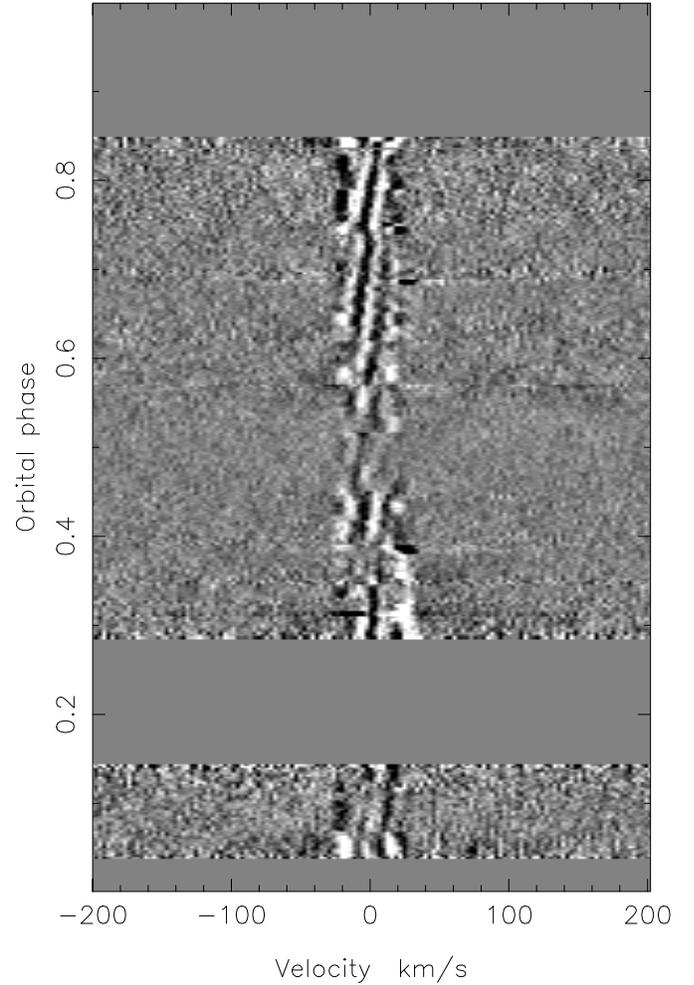}
\caption[]{Time series of deconvolved profiles derived from the
original WHT spectra, and secured over 17 nights observations in 1998,
1999 and 2000. The observations assume a grey albedo spectrum without
the addition of a synthetic planet signal. The greyscale runs from
black at $- 10^{-4}$ times the mean stellar continuum level, to white
at $+10^{-4}$. The velocity scale is in the reference frame of the
star.}
\label{fig:ph_noplanet}
\end{figure}

For any assumed axial inclination, the phase angle and line-of-sight
velocity are known at all times. The simulation procedure simply
consists of shifting and scaling the spectrum of HR 5694 according to
the orbit and phase function, co-multiplying it by an appropriate
geometric albedo spectrum, and adding it to the observed data. To
ensure a strong signal we used a simulated planet of radius 1.4
$\rmsub{R}{Jup}$ and wavelength-independent geometric albedo $p=0.5$,
which when viewed at zero phase angle should give a planet-to-star
flux ratio $\epsilon_{0}=f_p/f_*=0.98\times 10^{-4}$. We have chosen
a planetary radius greater than that expected by theory so as to
provide a simulated input signal strong enough to return an
unambiguous detection.

The resulting time series of deconvolved line profiles, shown in
Fig.~\ref{fig:ph_a0.5r1.4ax60}, demonstrates how the simulated planet
signal is recovered after the extraction process, with the planetary signal
clearly visible as a dark sinusoidal feature crossing from right
to left between phases 0.25 and 0.75. The weakening of the simulated
planetary signature near quadrature is caused mainly by the phase function.
The signal is further attenuated near quadrature by the
way in which the templates are computed: since the planet signature is
nearly stationary in this part of the orbit, some of the signal will
be removed along with the stellar profile if many observations are
made in this part of the orbit.

\begin{figure}
\psfig{figure=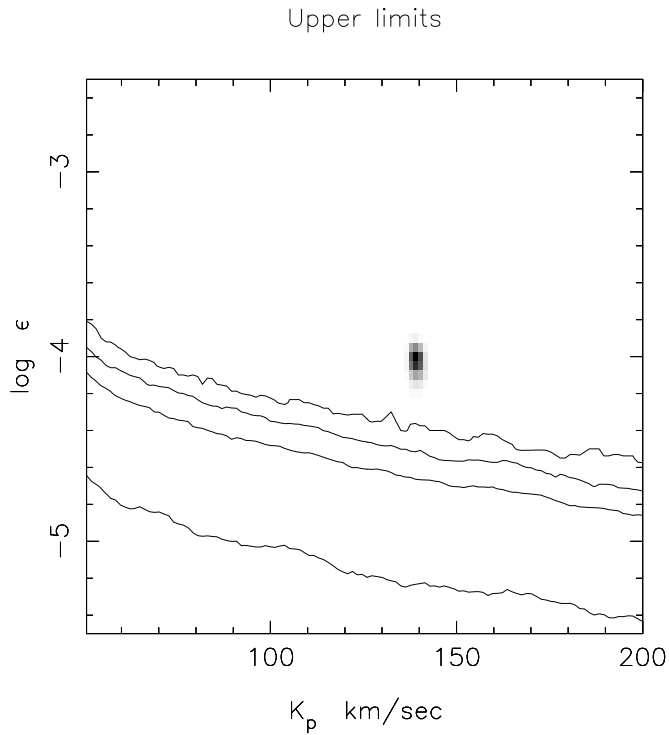,bbllx=106pt,bblly=73pt,bburx=481pt,bbury=404pt,angle=-90,width=8.6cm}
\caption[]{Relative probability map of model parameters $K_{p}$ and
$\log(\epsilon_{0})=\log p(R_{p}/a)^{2}$ for a simulated planet
signature with grey albedo $p=0.5$, $R_{p}=1.4 R_{Jup}$ and orbital
inclination of 60 $^\circ$. The contours show the confidence levels
at which candidate detections can be ruled out as being caused by
spurious alignments of non-Gaussian noise features. From top to
bottom, they show the 99.9\%, 99.0\%, 95.4\%\ and 68.4\%\ confidence
limits. The synthetic planet signature is detected well above the
99.9\%\ confidence limit.}
\label{fig:limits_a0.5r1.4ax60}
\end{figure}

\begin{figure}
\psfig{figure=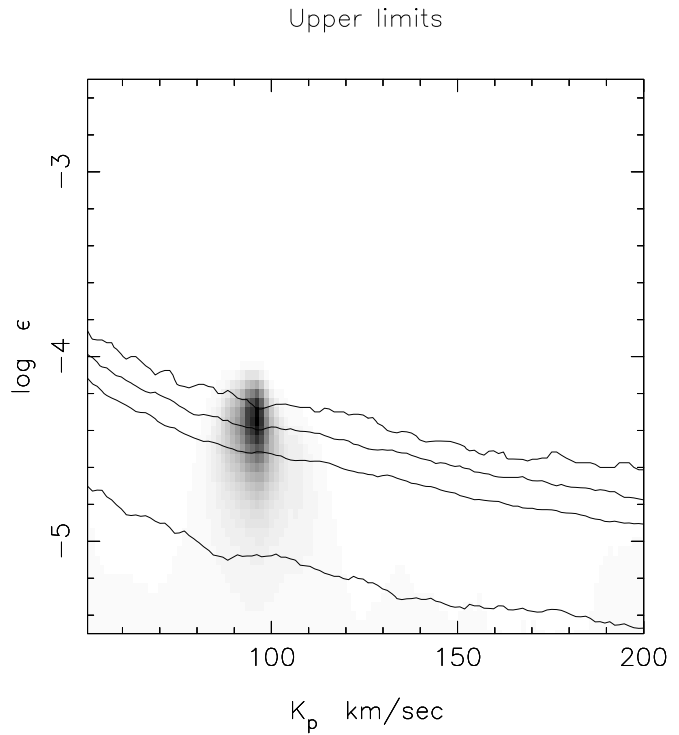,bbllx=106pt,bblly=73pt,bburx=481pt,bbury=404pt,angle=-90,width=8.6cm}
\caption[]{Relative probability map of model parameters $K_{p}$ and
$\log(\epsilon_{0})=\log p(R_{p}/a)^{2}$, derived from the WHT/UES
observations of $\tau$ Bootis, assuming a grey albedo spectrum. The
greyscale denotes the probability relative to the best-fit model,
increasing from 0 for white to 1 for black. A broad candidate feature
appears close to the 99.0\% confidence contour, with projected orbital
velocity amplitude $K_{p}= 97$~km~s$^{-1}$.  }
\label{fig:limits_noplanet}
\end{figure}

\begin{figure}
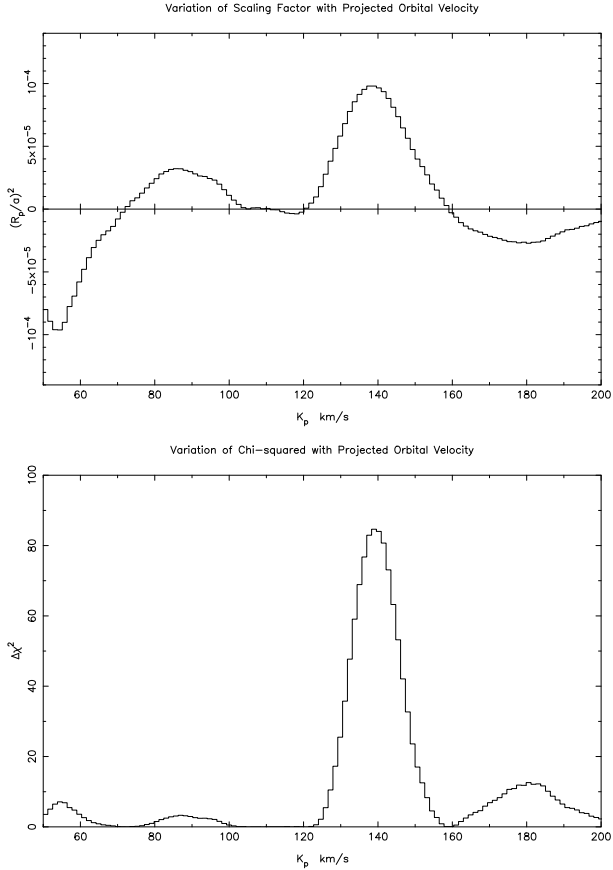

\psfig{figure=mc654_fig6.ps,angle=-90,width=8.6cm}
\psfig{figure=mc654_fig7.ps,angle=-90,width=8.6cm}
\caption{The upper panel shows the optimal scaling factor
$(R_{p}/a)^{2}$ plotted against orbital velocity amplitude $K_{p}$,
assuming a grey albedo spectrum for the simulated planet data. The
lower panel shows the associated reduction $\Delta\chi^{2}=84.6$, 
measured relative to the fit obtained in the absence of any
planet signal i.e. for $(R_{p}/a)^{2}$ = 0. Note that only positive
values of $(R_{p}/a)^{2}$ are physically plausible.}
\label{fig:mtomo_a0.5r1.4ax60}
\end{figure}

\begin{figure}
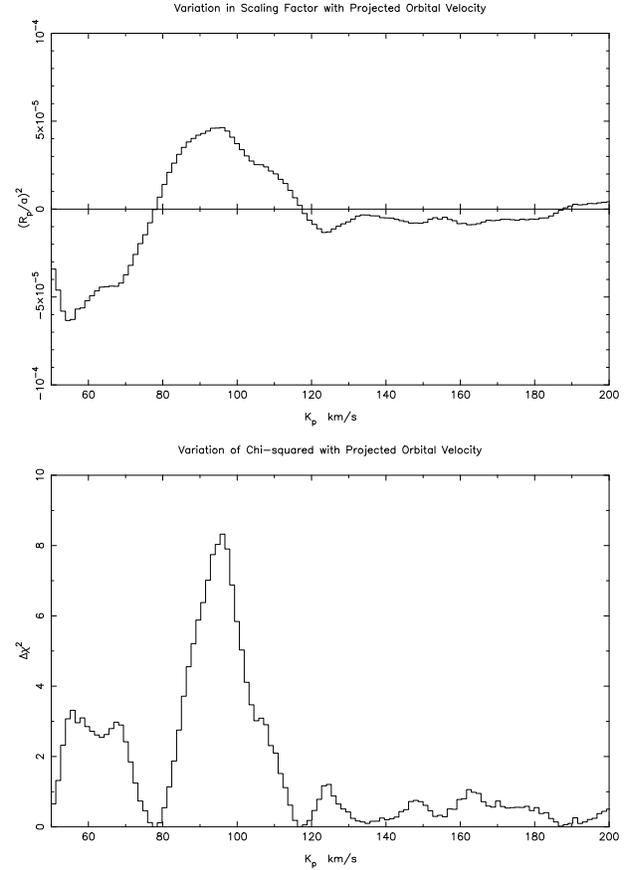

\psfig{figure=mc654_fig8.ps,angle=-90,width=8.6cm}
\psfig{figure=mc654_fig9.ps,angle=-90,width=8.6cm} \caption{As for
Fig.~\ref{fig:mtomo_a0.5r1.4ax60} but assuming a grey albedo spectrum
for the original data without a simulated planet signal. The lower
panel shows the associated reduction in $\Delta\chi^{2}=8.324$,
measured relative to the fit obtained in the absence of any planet
signal. The weak improvement in $\Delta\chi^{2}$ at
$K_{p}=60$~km~s$^{-1}$ corresponds to a negative value of
$(R_{p}/a)^{2}$ and is therefore physically implausible.  }
\label{fig:mtomo_noplanet}
\end{figure}

Figs.~\ref{fig:ph_a0.5r1.4ax60} and \ref{fig:ph_noplanet} both show a
``barber's-pole'' pattern of distortions in the residual stellar
profiles at low velocities. The phase variation in these undulations appears
consistent with sub-pixel shifts in the position of the spectra with respect
to the detector over the course of the night. Fortunately they only affect a
range of velocities at which the planet signature would in any case be
indistinguishable from that of the star.

The relative probabilities of the fits to the data for different
values of the free parameters $R_{p}/a$ and $K_{p}$ are given by
\begin{equation}
        P(K_{p},R_{p}/a)\propto\exp(-\chi^{2}/2),
\end{equation}
where
\begin{equation}
        \chi^{2}=\sum_{i,j}
        \frac{(D_{ij}-(R_{p}/a)^{2}H(v_{i},\phi_{j},K_{p}))^{2}}{\sigma^{2}_{ij}}.
\end{equation}
This is conveniently displayed in greyscale form as a function of
$K_{p}$ and $\log(\epsilon_{0})=\log p(R_{p}/a)^{2}$.  In
Fig.~\ref{fig:limits_a0.5r1.4ax60} we show the map for the simulated
observations, with the probabilities normalised to the most probable
value in the map.

The signal of the synthetic planet appears as a compact, dark feature
at $K_{p}=139$~km~s$^{-1}$ and $\log(\epsilon_{0})=-4.01$, i.e.
$\epsilon_{0}=0.98\times 10^{-4}$. This most probable combination of
orbital velocity and planet radius yields an improvement
$\Delta\chi^{2}=84.6$ with respect to the value obtained assuming no
planet is present (Fig.~\ref{fig:mtomo_a0.5r1.4ax60}).

To set an upper limit on the strength of the planet signal, or to
assess the likelihood that a candidate detection is spurious, we
need to compute the probability of obtaining such an improvement
in $\chi^{2}$ by chance alone. In principle this could be done using
the $\chi^{2}$ distribution for 2 degrees of freedom. In practice,
however, the distribution of pixel values in the deconvolved
difference profiles has extended non-Gaussian tails that demand a
more cautious approach.

Rather than relying solely on formal variances derived from photon
statistics, we use a ``bootstrap'' procedure to construct empirical
distributions for confidence testing, using the data themselves. In
each of 3000 trials, we randomize the order in which the 17 nights
of observations were secured, then we randomise the order in which the
observations were secured within each night. The re-ordered
observations are then associated with the original sequence of dates
and times. This ensures that any contiguous blocks of spectra
containing similar systematic errors remain together, but appear at a
new phase. Any genuine planet signal present in the data is, however,
completely scrambled in phase. The re-ordered data are therefore as
capable as the original data of producing spurious detections through
chance alignments of blocks of systematic errors along a single
sinusoidal path through the data. We record the least-squares
estimates of $\log(\epsilon_{0})=\log p(R_{p}/a)^{2}$ and the associated
values of $\chi^{2}$ as functions of $K_{p}$ in each trial.

The percentage points of the resulting bootstrap distribution are
shown as contours in Figs.~\ref{fig:limits_a0.5r1.4ax60} and
\ref{fig:limits_noplanet}. From bottom to top, these contours give
the 68.4\%\, 95.4\%\, 99.0\%\ and 99.9\%\ bootstrap upper limits on
the strength of the planet signal. The 99.9\%\ contour, for example,
represents the value of $\log(\epsilon_{0})$ that was only exceeded in
3 of the 3000 trials at each $K_{p}$.

\section{Results and Discussion}
\label{sec:results}

The results of this analysis appear on the relative probability map of
model parameters $K_{p}$ and $\log(\epsilon_{0})=\log p(R_{p}/a)^{2}$,
shown at Fig.~\ref{fig:limits_noplanet}. The calibrated confidence
levels allow us to achieve our primary aim of constraining the radius
and albedo of the planet.  However, there exists a significant
candidate feature close to the 99\%\ level that requires further
investigation. In the subsequent discussion, we therefore also
explore the possibility that this feature could represent a genuine
planetary detection.

If the feature were genuine, the projected orbital velocity amplitude
$K_{p}\simeq 97~(\pm 10)~km~s^{-1}$ yields an orbital inclination of 37
($\pm 5$)$^\circ$. This would be consistent with the star's rotation
being tidally locked to the planet's orbit and implies a mass for
$\tau$ Bootis b of $M_{p} = 7.28(\pm 0.83) M_{Jup}$.

We emphasise that, although the feature appears very close to the peak
of the prior probability distribution projected onto $K_{p}$, shown in
Fig.~\ref{fig:prior}, there remains a distinct possibility that the
candidate detection is a consequence of spurious noise and as such we
should proceed with caution.

\subsection{Upper Limits on Grey Albedo}
\label{sec:albedo_lims}

The grey albedo model assumes that at all times the planet-star flux
ratio is independent of wavelength. For an assumed planetary radius
we can thus use Equation~\ref{eq:eps0} to constrain the geometric
albedo. Table~\ref{tab:greyresults} lists the upper limits on the
albedo at various levels of significance, for the planetary radius
$R_{p}=1.2 R_{Jup}$ predicted by current theoretical models
\cite{guillot96,burrows2000}.

\begin{table}
\caption[]{Upper limits on the grey albedo for the atmosphere of
$\tau$ Bootis b assuming a radius of 1.2 $R_{Jup}$. The upper limits
are quoted for an assumed $K_{p} \simeq 94$~km~s$^{-1}$, at the peak
of the prior probability distribution for $K_{p}$.}
\begin{tabular}{ccc}
\hline \\
False Alarm 	& $p\left(R_{p}/a\right)^{2}$	& Upper Albedo Limit	\\
Probability	&				& $p$			\\
\\
0.1 \%\		& 0.561 E-04			& 0.39		\\
1.0 \%\		& 0.403 E-04			& 0.28		\\
4.6 \%\		& 0.305 E-04			& 0.21		\\
\\
\hline \\
\end{tabular}
\label{tab:greyresults}
\end{table}

The contours in Fig.~\ref{fig:limits_noplanet} produced by the
bootstrap simulation constrain the maximum reflected flux ratio at
opposition to be $\epsilon_{0} \leq 0.561 \times 10^{-4}$ at the 99.9 \%
confidence level, assuming a projected orbital velocity $K_{p}$ at the
peak of the prior probability distribution, i.e. $K_{p} \simeq
94$~km~s$^{-1}$. This would limit the geometric albedo of the planet
to $p\leq0.39$. We note that this is a similar result to that
obtained by \scite{charb99} at the same inclination. Both studies
assume a grey albedo, $R_{p} = 1.2 R_{Jup}$, synchronous rotation of
the star and hence a reflected version of the stellar spectrum with no
rotational broadening. The candidate feature that appears in
Fig.~\ref{fig:limits_noplanet} would, if genuine, yield a grey
geometric albedo of $p~=~0.32~(\pm0.13)$ for a planet of this radius.

\subsection{Upper Limits on Radius}
\label{sec:radius_lims}

\begin{figure}
\psfig{figure=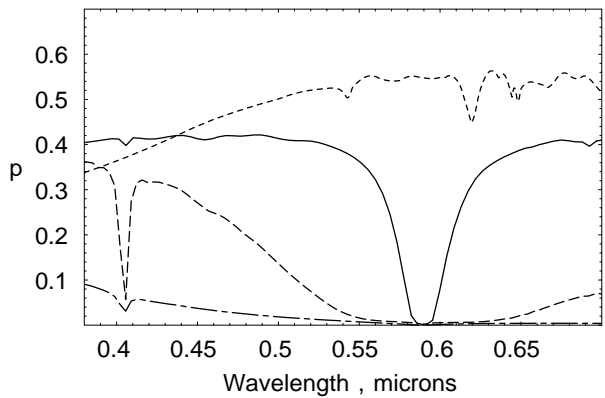,width=8.5cm}
\caption[]{Geometric albedo spectra for the Class V (solid line),
isolated (dashed) and irradiated (dot-dashed) Class IV models of
\scite{sudarsky2000}, plotted over the wavelength range we
observed. Together with a grey model of albedo $p=0.3$, the Class V and
isolated Class IV models were used to probe the wavelength dependence
of candidate reflected-light signals. The geometric albedo spectrum
of Jupiter (short dashes) is shown for comparison.}
\label{fig:albsgeom}
\end{figure}

Here we investigate how atmospheric albedo models can be incorporated
into the signal analysis to place upper limits on the size of the
planet. A non-grey albedo model is built into the formation of the
least-squares deconvolved profile, by scaling the strengths of the
lines in the deconvolution mask by a factor equal to the geometric
albedo at each line's wavelength. The scale factor produced by the
matched-filter analysis is then directly proportional to
$(R_{p}/a)^{2}$ (see Equation~\ref{eq:eps0}) and is calibrated by
injecting the signature of a planet of known radius with the specified
albedo spectrum into the data. The method is described in detail by
\scite{cameron02}.

The theoretical models we consider here are those proposed by
\scite{sudarsky2000} for a range of extrasolar giant planets, as shown
at Fig.~\ref{fig:albsgeom}. These models are grouped primarily by
their mass and orbital distance from the host star, factors which in
turn influence their effective surface temperature, surface gravity
and hence radius. We recognise, however, that recent observations of
the atmosphere of HD 209458b \cite{charb2002} suggest there may be
less sodium absorption than predicted in these and other models
\cite{brown2001,hubbard2001}.

\begin{table}
\caption[]{Upper limits on planet radius for various albedo models. 
The limits are quoted for an assumed $K_{p} \simeq 94$~km~s$^{-1}$, at
the peak of the prior probability distribution for $K_{p}$. Note that
the results for the grey albedo model are given for a geometric albedo
of $p=0.3$.  Radii for other grey model albedos can be obtained by
dividing Column 4 by $\sqrt{p/0.3}$ .}
\begin{tabular}{cccc}
\hline \\
Albedo Model    & False Alarm		& $(R_{p}/a)^{2}$   	& $R_{p}/R_{Jup}$	\\
                & Probability		&                       & Upper Limit		\\
\\
Grey		& 0.1\%\		& 1.87 E-04             & 1.37                  \\
($p = 0.3$)	& 1.0\%\		& 1.34 E-04		& 1.16			\\
		& 4.6\%\		& 1.02 E-04		& 1.01			\\
\\
Class V		& 0.1\%\                & 1.16 E-04		& 1.08			\\
		& 1.0\%\                & 0.91 E-04		& 0.95			\\
                & 4.6\%\                & 0.63 E-04		& 0.79			\\
\\
Class IV	& 0.1\%\                & 1.50 E-04		& 1.22			\\
(Isolated)	& 1.0\%\                & 1.13 E-04		& 1.06			\\
                & 4.6\%\                & 0.87 E-04		& 0.93			\\
\\
\hline \\
\end{tabular}
\label{tab:rad_results}
\end{table}

\begin{table*}
\caption[]{Projected orbital velocity peak, planet radius,
$\Delta\chi^{2}$ and false-alarm probabilities (FAP) for the candidate
feature, on the basis that it represents a genuine detection.The
second FAP weights $K_{p}$ in proportion to the prior probability
density distribution (Fig.~\ref{fig:prior}).}
\begin{tabular}{cccccc}
\hline \\
Albedo Model	& $K_{p}$	& $R_{p}/R_{Jup}$ & $\Delta\chi^{2}$	& FAP	           & FAP	     \\
		& ($km~s^{-1}$)	&		  &	                & (Uniform Weight) & ($K_{p}$ Prior) \\
\\
Grey ($p=0.3$)	& 97 ($\pm 8$)	& $1.24\pm 0.25$	& 8.324		& 0.147	& 0.036			\\
\\
Class V		& 95 ($\pm 8$)	& $1.08\pm 0.19$	& 12.06 	& 0.032	& 0.003			\\
\\
Class IV	& 90 ($\pm 8$)	& $1.18\pm 0.20$	& 9.227		& 0.092	& 0.032			\\
\\
\hline \\
\end{tabular}
\label{tab:fap_results}
\end{table*}

\subsubsection{Grey Albedo Model}
\label{sec:grey}

At the most probable values in the prior distribution ($K_{p} \simeq
94$~km~s$^{-1}$), assuming a grey albedo model of $p=0.3$, the
0.1\%, 1.0\% and 4.6\% upper limits on the planet/star flux ratio
$\epsilon_{0}$ correspond to upper limits on the planet radius, as
detailed in Table~\ref{tab:rad_results}. With a higher assumed
geometric albedo, the planet's radius is more strongly constrained.

Our potential planet signal yields an improvement $\Delta\chi^{2} =
8.324$ over the model fit obtained assuming no planet signal is
present (Fig.~\ref{fig:mtomo_noplanet}). We used the bootstrap
simulations to determine the probability that a spurious feature with
$\Delta\chi^{2} > 8.324$ could be produced by a chance alignment of
noise features in the absence of a genuine planet signal. It is
important to note that the bootstrap contours only give the
false-alarm probability if the value of $K_{p}$ is known in advance,
which is not the case here. The true false-alarm probability is
greater, being the fraction of bootstrap trials where spurious peaks
at any plausible value of $K_{P}$ can exceed the $\Delta\chi^{2}$ of
the candidate. If we assume that all values of $K_{P}$ are equally
likely in the range 50~km~s$^{-1} < K_{P} < 162$~km~s$^{-1}$, the
false-alarm probability is found to be 14.7\%\, via the method
described more comprehensively in \scite{cameron02} Appendix E.

In practice, however, we are more likely to believe that a feature
detected near the peak of the prior probability distribution for
$K_{p}$ is genuine, than if the feature appeared at a velocity that
was physically implausible given our existing knowledge of the system
parameters.  We can therefore use our prior estimation of $K_{P}$ to
weight the false-alarm probabilty, in the manner discussed in
\scite{cameron02} Appendix E. We find from Table~\ref{tab:fap_results}
that the false-alarm probability drops to 3.6\% when prior knowledge
of $K_{p}$ is accounted for. For comparison, we find that a matched
filter analysis of the simulated planet data
(Fig.~\ref{fig:mtomo_a0.5r1.4ax60}) sees an improvement of
$\Delta\chi^{2} = 84.6$ above the value obtained assuming no planet
signal is present. This is far greater than the $\Delta\chi^{2} =
42.3$ produced at any $K_{p}$ in the bootstrap trials. The
false-alarm probability is therefore substantially less than one part
in 3000, and as such the ``detection'' of the simulated signal is
secure.

\subsubsection{Class V Model}
\label{sec:classv}

\begin{figure}
\psfig{figure=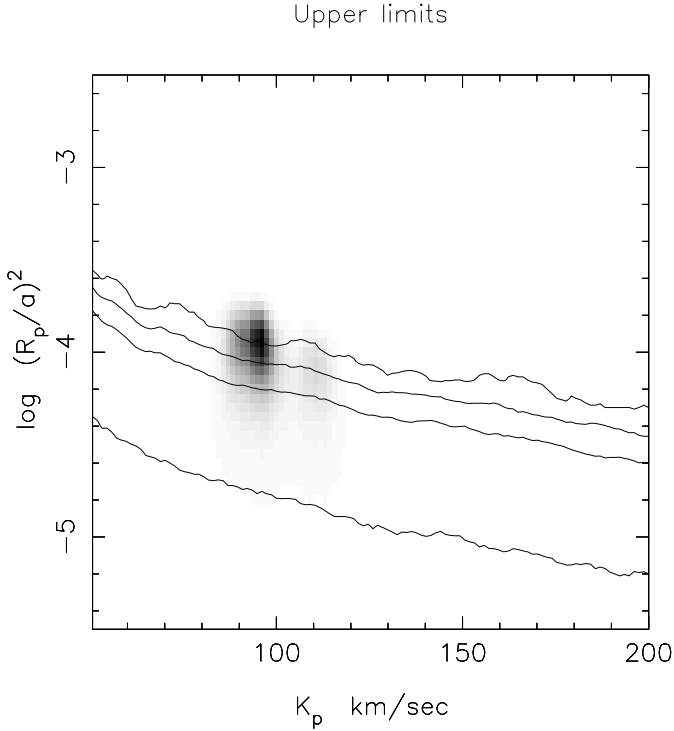,bbllx=106pt,bblly=73pt,bburx=481pt,bbury=404pt,angle=-90,width=8.6cm}
\caption{Relative probability map of model parameters $K_{p}$ and
$\log(R_{p}/a)^{2}$, derived from the WHT/UES observations of $\tau$
Bootis, assuming the albedo spectrum to be that of a Class V roaster. 
The greyscale and contours are defined as in
Fig.~\ref{fig:limits_a0.5r1.4ax60}.  }
\label{fig:limits_classv}
\end{figure}

\begin{figure}
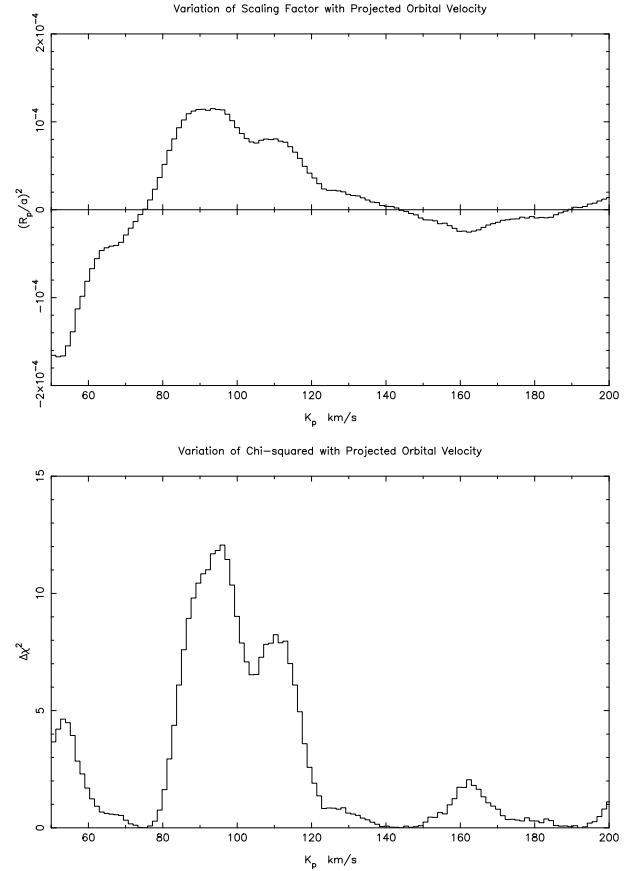

\psfig{figure=mc654_fig12.ps,angle=-90,width=8.6cm}
\psfig{figure=mc654_fig13.ps,angle=-90,width=8.6cm}
\caption{As for Fig.~\ref{fig:mtomo_a0.5r1.4ax60} but assuming a Class V
spectrum for the original data without a simulated planet signal.The lower
panel shows the associated reduction in $\Delta\chi^{2}$ of 12.06, measured
relative to the fit obtained in the absence of any planet signal.}
\label{fig:mtomo_classv}
\end{figure}

The ``Class V roaster'' is the most highly reflective of the models
published by \scite{sudarsky2000}. It is characteristic of planets
with $T_{eff}\ge 1500$ K and/or surface gravities lower than $\sim 10$
m s$^{-2}$, and as such is associated with lower mass planets, such as
$\upsilon$ And b. The model predicts a silicate cloud deck located
high enough in the atmosphere that the overlying column density of
gaseous alkali metals is low, allowing a substantial fraction of
incoming photons at most optical wavelengths to be scattered back into
space. There remains, however, a substantial absorption feature
around the Na I D lines, as shown in Fig.~\ref{fig:albsgeom}.

We carried out the deconvolution using the same line list as for the
grey model, but with the line strengths attenuated using the Class V
albedo spectrum (see \scite{cameron02} Appendix C). We calibrated the signal
strength as described in \scite{cameron02} Appendix D, by
injecting an artificial planet signature consisting of the spectrum of
HR 5694, attenuated by the Class V albedo spectrum and scaled
to the signal strength expected for a planet with $R_{p}=1.4 R_{Jup}$.

The form of the Class V probability map, as shown in
Fig.~\ref{fig:limits_classv} is similar to that encountered for the
grey albedo spectrum. The resulting upper limits on the planet radius
are detailed at Table~\ref{tab:rad_results}, with the corresponding
false-alarm probabilities listed in Table~\ref{tab:fap_results}. We
find the feature produces a local probability maximum near $K_{p} =
95$~km~s$^{-1}$, with an improvement in $\Delta\chi^{2}$ over the grey
albedo model of 12.06, as plotted in Fig.~\ref{fig:mtomo_classv}. 
This improvement translates to a reduced FAP (unweighted) of 3.2\%,
however, the position of the best-fitting $K_{p}$ matches the prior
probability maximum of $K_{p} = 94$~km~s$^{-1}$ so closely that the
overall FAP is 0.3\%, substantially lower than that obtained for the
grey albedo case. This suggests strongly that the features in the
data that give rise to this signal originate predominantly at blue
wavelengths. If the candidate feature we observe were genuine, it
would indicate a Class V planet of radius $1.08 (\pm 0.19)~R_{J}$,
which is in line with with current theory
\cite{guillot96,burrows2000}.

\subsubsection{Isolated Class IV model}
\label{classiv}

\begin{figure}
\psfig{figure=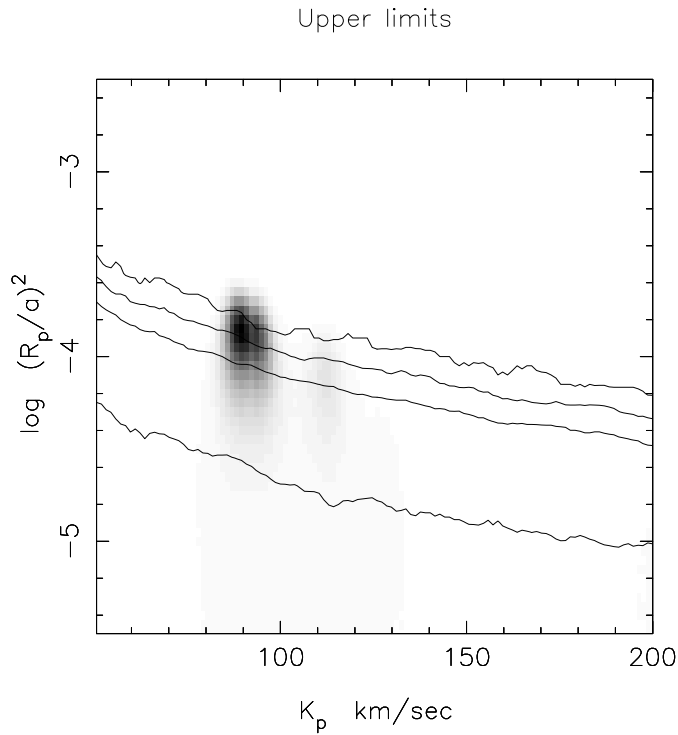,bbllx=106pt,bblly=73pt,bburx=481pt,bbury=404pt,angle=-90,width=8.6cm}
\caption{Relative probability map of model parameters $K_{p}$ and
$\log(R_{p}/a)^{2}$, derived from the WHT/UES observations of $\tau$
Bootis, assuming the albedo spectrum to be that of an ``isolated''
Class IV gas giant. The candidate feature, if genuine, corresponds to
the detection of a $1.18 (\pm 0.20)~R_{J}$ planet. The greyscale and
contours are defined as in Fig.~\ref{fig:limits_a0.5r1.4ax60}.  }
\label{fig:limits_classiv}
\end{figure}

\begin{figure}
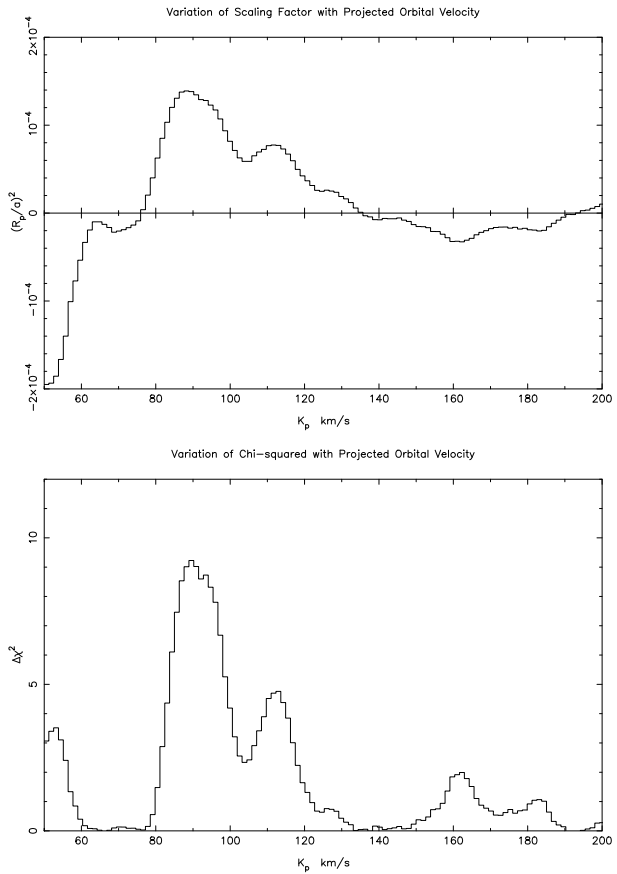

\psfig{figure=mc654_fig15.ps,angle=-90,width=8.6cm}
\psfig{figure=mc654_fig16.ps,angle=-90,width=8.6cm} \caption{As for
Fig.~\ref{fig:mtomo_a0.5r1.4ax60} but assuming a Class IV ``isolated''
spectrum for the original data without a simulated planet signal. The
lower panel shows the associated reduction in $\Delta\chi^{2}$ of
9.227, measured relative to the fit obtained in the absence of any
planet signal.}
\label{fig:mtomo_classiv}
\end{figure}

The ``Class IV'' models of \scite{sudarsky2000} have a more
deeply-buried cloud deck than the Class V models and are probably more
closely applicable to $\tau$ Boo b given its relatively high surface
gravity.  The resonance lines of Na I and K I are strongly saturated,
with broad damping wings due to collisions with H$_{2}$ extending over
much of the optical spectrum (Fig.~\ref{fig:albsgeom}).

We used the procedures described above to deconvolve and back-project
the data assuming an ``isolated'' Class IV spectrum. Although this
model does not take full account of the effects of irradiation of the
atmospheric temperature-pressure structure, it is a useful compromise
between the Class V models and the very low albedos found with
irradiated Class IV models. The resulting time-series of deconvolved
spectra is noisier than the Class V and grey-albedo versions,
because lines redward of 500 nm contribute little to the deconvolution.

The probability map (Fig.\ref{fig:limits_classiv}) derived from the
bootstrap matched-filter analysis again shows a marginally significant
candidate reflected-light feature, but for this albedo model the peak
of the distribution is shifted to $K_{p}=90$~km~s$^{-1}$. The fit to
the data is slightly better than in the grey albedo case, giving an
improvement of $\Delta\chi^{2}=9.227$ over the no-planet hypothesis.
The bootstrap analysis returns an unweighted false-alarm probability
of 9.2\%, but the displacement of the most probable value of $K_{p}$
(90~km~s$^{-1}$) away from the prior probability maximum at $K_{p} =
94$~km~s$^{-1}$ means the overall Bayesian FAP is 3.2\%, slightly
lower than in the grey albedo case.

The increase in noise associated with the extraction of the Class IV
simulation produces more loosely constrained upper limits on the
planetary radius, as set out in Table~\ref{tab:rad_results}. We note
that if the candidate feature were genuine, it would indicate a Class
IV planet of radius $1.18 (\pm 0.20)~R_{J}$, still in line with
current expectations \cite{guillot96,burrows2000}.

\section{Conclusion}
\label{sec:conclusion}

We have re-analysed the WHT echelle spectra obtained for the F7V star
$\tau$ Bootis during 1998, 1999 and 2000. By assuming that

(a) the rotation of $\tau$ Bootis is tidally locked to the orbit of
the planetary companion, suggesting a orbital inclination of $i \sim
40^\circ$, and

(b) the planet radius $R_{p} \simeq 1.2 M_{Jup}$, in accordance with
the general theoretical predictions of \scite{guillot96,burrows2000}
and with observations of the planetary transits across HD 209458
\cite{charb2000},

\smallskip
\noindent we are able to rule out a reflective planet with a grey
albedo greater than $p=0.39$ to the 99.9\% confidence level. The
alternative approach of adopting the specific grey ($p=0.3$), Class V
and Class IV albedo models of \scite{sudarsky2000} places
model-dependent upper limits on the planetary radius. The results
indicate with 99.9\%\ confidence that the upper limits are
$1.37~R_{Jup}$, $1.08~R_{Jup}$ and $1.22~R_{Jup}$ respectively for the
three albedo models considered.

Our analysis reveals a candidate signal of marginal significance with
a projected orbital velocity amplitude of $K_{p}~=~97$~km~s$^{-1}$,
assuming a grey albedo spectrum. If genuine, this would suggest an
orbital inclination close to $\sim 37^\circ$, a planet mass $M_{p} =
7.28~(\pm 0.83)~M_{Jup}$ and a grey geometric albedo of $p=0.32~(\pm
0.13)$, assuming $R_{p} = 1.2~R_{Jup}$. If we feign complete
ignorance of the value of $K_{p}$, our bootstrap Monte Carlo
simulations give a probability ranging from 3 to 15\% that the
detected feature is a consequence of spurious noise from the analysis. 
When taking into account our prior knowledge of the system these false
alarm probabilities drop to below 3\%.  

In particular, the Class V albedo model -- in which only the spectrum
shortward of 550 nm is unaffected by Na I D absorption -- gives a
false-alarm probability of only 0.3\%\ when the prior probability
distribution for $K_{p}$ is taken into account using Bayes' Theorem. 
However, we consider this is still too large an uncertainty for us to
claim a bona fide detection. Our simulations show that a
statistically unassailable detection should produce a clearly visible,
dark streak along the planet's trajectory in the trailed spectrogram,
and no such streak is apparent even for the Class V model.

The observations in the 2000 season were conducted at orbital phases
optimised to produce the strongest possible signal at a star-planet
separation in velocity space sufficient to avoid blending problems. 
By adopting similar observing strategies on 8m-class telescopes,
future reflected light searches of $\tau$ Bootis should be able to
double the effective planetary signal contained in the WHT data
described here, in only a small fraction of the 17 nights devoted to
this search. Indeed, 2 optimally-phased clear nights on the
KeckI/HIRES combination should be able to reproduce our results,
whilst the increased efficiency of the HDS spectrograph would allow
Subaru to achieve very close to this depth of search over the same
timescale. With this in mind, we believe that $\tau$ Bootis remains a
suitable target for future reflected light searches on 8m-class
telescopes.

\newpage

\bibliography{mc654} \bibliographystyle{mn}

\section*{Acknowledgements}

This work is based on observations made with the William Herschel
Telescope, operated on the island of La Palma by the Isaac Newton
Group in the Spanish Observatorio del Roque de los Muchachos of the
Instituto de Astrofisica de Canarias. The initial data reduction was
carried out using the ECHOMOP and FIGARO software supported by the
Starlink Project, on PC/Linux hardware funded through a PPARC rolling
grant. ACC and KDH acknowledge the support of PPARC Senior
Fellowships during the course of this work.

We thank David Sudarsky and Adam Burrows for providing us with
listings of their Class IV and Class V albedo models. We also thank
Geoff Marcy for his updates on the orbital ephemeris of $\tau$ Boo b.

\newpage

\appendix

 
\section{Calibrating the Matched-filter analysis}
\label{sec:calibration}

The purpose of incorporating a simulated planet signature into our
analysis is two-fold. First, it allows us to ensure that any
planetary signal, real or simulated, is maintained through the
template subtraction and deconvolution procedures and can be recovered
during the subsequent matched filter analysis. In doing so we can
measure the degree to which any simulated signal is attenuated and
infer that any real planetary signal would suffer a similar fate. 
Second, by using a suitable calibration factor, we can ensure that the
matched filter detection for the simulated planet, appears at the
expected position in the resulting
$\log(\epsilon_{0})=\log~p(R_{p}/a)^{2}$ vs $K_{p}$ probability map,
i.e. at $\log(\epsilon_{0})=-4.01$ in
Fig.~\ref{fig:limits_a0.5r1.4ax60}. Thus any potential detections
within the real data (Fig.~\ref{fig:limits_noplanet}) would be
suitably compensated for losses imposed by the extraction and analysis
procedures.

We model the reflected-light signal as a time sequence of Gaussians
with appropriate velocities and relative amplitudes according to
\begin{eqnarray}
G(v,\phi,K_{p}) &=& \frac{W_{*}}{\Delta v_{p}\sqrt{\pi}}
g(\phi,i)\times\nonumber\\
&&\times\exp\left[-\frac{1}{2}
\left(\frac{v-K_{p}\sin\phi}{\Delta v_{p}}\right)^{2}\right].
\label{eq:basfunc}
\end{eqnarray}
where the amplitude $K_{p}$ of the sinusoidal velocity variation is
determined by the system inclination and stellar mass.

\begin{figure}
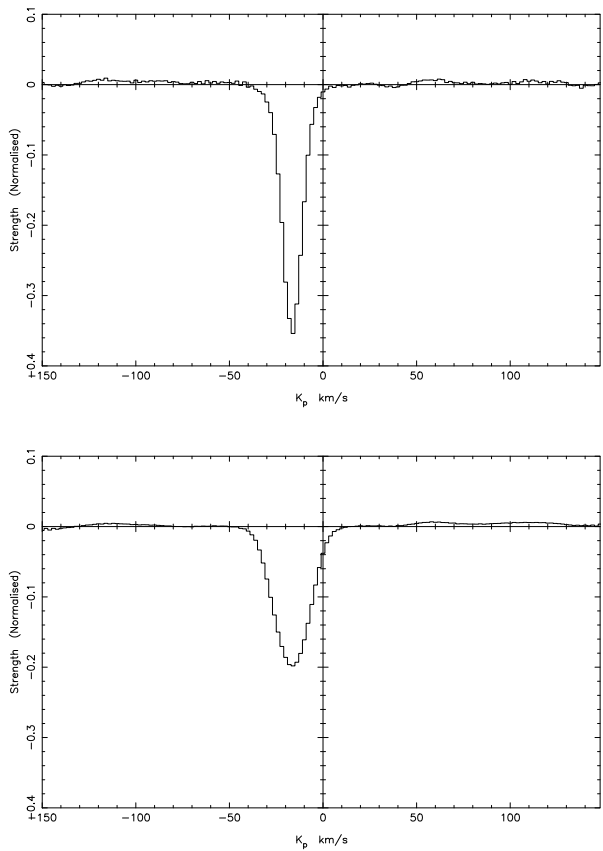

\psfig{figure=mc654_fig17.ps,angle=-90,width=8.6cm}
\psfig{figure=mc654_fig18.ps,angle=-90,width=8.6cm}
\caption{Deconvolved line profiles for HR 5694 and $\tau$ Bootis,
showing the average broadening exhibited by all lines within their
echelle spectra. The line profiles exhibit significantly different
shapes due to the difference between the two stars' rotation rates. 
The normalised equivalent widths of the line profiles are $W_{*}$ =
4.45 and $W_{*}$ = 4.42 respectively.}
\label{fig:line_profiles}
\end{figure}

The variable factor we use to calibrate the strength of the detected
signal is the equivalent width ($W_{*}$) of the stellar component of
the composite line profile. By deconvolving the observed spectrum of
HR 5694 with a list of the relative strengths of its spectral lines,
we obtain a composite line profile exhibiting the broadening function
that is representative of all the lines recorded in the spectrum. The
deconvolved line profile is shown at Fig.~\ref{fig:line_profiles}
alongside that for $\tau$ Bootis itself. We recall that HR 5694 was
chosen to best mimic the non rotationally broadened line profiles
reflected from the near tidally locked planet.

The planet signal should take the form of a faint copy of the stellar
spectrum, as sharp as the deconvolved profile of HR 5694, located deep
within the noise of the composite deconvolved residual profile of
$\tau$ Bootis, i.e. the deconvolved profile of the residual $\tau$
Bootis spectrum following stellar subtraction.

In effect the matched-filter analysis compares the $W_{*}$ value with
the strength of the best fit Gaussian filter from the phased residual
profiles, as at Fig.~\ref{fig:ph_a0.5r1.4ax60}. Ideally, by using
$W_{*}$ = 4.45, any simulated planet signal should be recovered at the
correct level within the $\log(\epsilon_{0})$ vs $K_{p}$ probability
map. However, analysis of the dataset following injection of a
synthetic planet signal showed a 15\% reduction in the equivalent
width ($W_{*}$ = 3.85) was required to recover the fake grey albedo
planet's signal at the correct strength. Any reduction in the
strength of the simulated signal during the various processes would
indeed manifest itself as a reduction in the $\log(\epsilon_{0})$
scaling factor, and would thus appear fainter than expected,
necessitating a correction to $W_{*}$. It is found that each of the
extraction processes contributes to the signal reduction, with the PCA
fixed noise removal (\scite{cameron02} Appendix B) contributing 9\% to
the total loss. Any genuine reflected light signal should undergo a
similar loss. In calibrating the grey albedo model
(Section~\ref{sec:grey}) we have therefore needed to correct for
this signal loss during the extraction process. Similar, but less
significant corrections ($\sim$ 10\%) had to be applied to the Class IV
and Class V models in order to calibrate the $(R_{p}/a)^{2}$ values
produced by the matched-filter analysis.

\end{document}